\documentclass[12pt]{article}
\begin{document}
\baselineskip=15.5pt
\renewcommand{\theequation}{\arabic{section}.\arabic{equation}}
\pagestyle{plain}
\setcounter{page}{1}
\begin{titlepage}

\leftline{\tt hep-th/0201149}

\vskip -.8cm

\rightline{\small{\tt CALT-68-2371}}
\rightline{\small{\tt CITUSC/02-002}}

\begin{center}

\vskip 2 cm

{\Large {The Equality of Solutions in Vacuum String Field Theory}}

\vskip 1.5cm
{\large Takuya Okuda}

\vskip 1.5cm

{California Institute of Technology 452-48, Pasadena, CA 91125}

\smallskip
{\tt takuya@theory.caltech.edu}

\vskip 2cm

{\bf Abstract}
\end{center}

We analytically prove that the matter solution of vacuum string field
theory constructed by Kostelecky and Potting is the matter sliver state.
We also give an analytical proof that the ghost solution by Hata and Kawano
is the sliver state in the twisted ghost CFT.
It is also proved that the candidate state for the tachyon proposed by Hata and
Kawano can be identified with the state constructed by Rastelli, Sen and Zwiebach
using CFT.
Our proofs are based on the techniques recently developed by Okuyama.
\end{titlepage}

\newpage


\section{Introduction}
\setcounter{equation}{0}
In the development of vacuum string field theory (VSFT)
\cite{Rastelli:2000hv}-\cite{Hata-Kawano},
two basic approaches have been taken, namely,
the operator formalism and the CFT formalism.
In the operator formalism, computations can be done explicitly and algebraically.
The formalism involves, however, infinite dimensional matrices and their determinants,
making it necessary to rely on numerical tests using level truncation.
On the other hand, the CFT formalism gives a geometrical
picture to various aspects of VSFT
and often analytical computations are possible,
but the techniques are more abstract.
With advantages and disadvantages for each method present,
both approaches have been employed to deal with questions in VSFT.

Consider, for example, the equation of motion
\begin{equation}
{\cal Q} \Psi + \Psi \ast \Psi = 0
\end{equation}
of VSFT. Since the kinetic operator $\cal{Q}$ consists of ghost
operators only, for a state of the form $\Psi^{(m)}\otimes
\Psi^{(g)}$, the equation of motion can be separated into the
matter part and the ghost part:
\begin{equation}
\Psi^{(m)} \ast \Psi^{(m)} =\Psi^{(m)},
{\cal Q}\Psi^{(g)}+\Psi^{(g)}\ast \Psi^{(g)}=0.
\end{equation}

For the matter part, Kostelecky and Potting
constructed a solution $\Psi_m$
in the operator language \cite{Kostelecky-Potting}.
Rastelli, Sen and Zwiebach suggested \cite{RSZclassical} that
this solution can be identified with the matter sliver state $\Xi_m$
which is defined as a surface state in the matter CFT,
giving numerical evidence based on level truncation.

To have a solution for the ghost part of the equation of motion,
the kinetic operator ${\cal Q}$ must be fixed.
Hata and Kawano showed that on some reasonable assumptions,
the existence of a nontrivial solution in the Siegel gauge
determines the kinetic operator ${\cal Q}$.
They then constructed a solution $\Psi_g$ and its corresponding kinetic
operator in the operator formalism \cite{Hata-Kawano}.
Gaiotto, Rastelli, Sen, and Zwiebach proposed
\cite{GRSZstructure} \cite{RSZnote}
that the kinetic operator ${\cal Q}$ is a mid-point
$c$ ghost insertion, and gave a ghost solution for
that ${\cal Q}$ as the sliver state
${\Xi'_g}$ in the twisted ghost CFT.
They further proposed that their ghost operator and solution can be
identified with those of Hata and Kawano and confirmed this identification
by level truncation analysis.

Following the spectrum analysis of the star algebra by Rastelli,
Sen, and Zwiebach \cite{RSZspectroscopy},
Okuyama  developed methods \cite{Okuyama-kinetic}
to analytically prove the equality
of the kinetic operators proposed by Hata and Kawano and that defined by
Gaiotto, Rastelli, Sen, and Zwiebach.

In this paper, we will use Okuyama's methods to analytically prove
the equality of the solution in the operator formalism and that in
the CFT formalism. In other words, we will analytically show that
the matter solution $\Psi_m$ by Kostelecky and Potting is the
matter sliver state $\Xi_m$ and also that the ghost solution
$\Psi_g$ by Hata and Kawano is the sliver state $\Xi'_g$ of the
twisted ghost CFT. Also, it is proved that the candidate tachyon
state proposed by Hata and Kawano is correctly reproduced by the
CFT construction as claimed by Rastelli, Sen, and Zwiebach \cite{RSZnote}.

The outline of this paper is the following.
In section \ref{def}, we review the methods of Okuyama
\cite{Okuyama-kinetic} \cite{Okuyama-tension}.
In section \ref{proof}, we give  analytical proofs
that the solutions and the candidate tachyon states in the operator
and CFT formalisms are identical.

\section{Review of Methods}\label{def}
\setcounter{equation}{0}
In this section, we review Okuyama's methods \cite{Okuyama-kinetic}
and set conventions.
The zero momentum sector of the 3-string vertex is given by \cite{Gross-Jevicki-1}
\cite{Gross:1986fk}
\begin{equation}
|V_3 \rangle=\exp\left(
\sum_{r,s=1}^{3}\sum_{n,m=1}^{\infty}\left[
-\frac{1}{2}a_{n\mu}^{(r)\dagger}V_{nm}^{rs}a_m^{(s)\mu\dagger}
+c_{-n}^{(r)}\widetilde{V}_{nm}^{rs}b_{-m}^{(s)}+c_{-n}^{(r)}\widetilde{v}_n^{rs}
b_0^{(s)}
\right]\right)
\bigotimes_{r=1}^{3}c_0^{(r)}c_1^{(r)}|0\rangle_r.
\end{equation}
We are using the convention $\alpha '=1$.
We write $M=CV^{11}$,
where $C_{nm}=(-1)^n\delta_{nm}$ is the twist matrix.
In \cite{RSZspectroscopy}, the eigenvectors and the eigenvalues of $M$ were
identified.
The eigenvector $v^{(k)}$
with eigenvalue $k\in {\bf R}$ is defined by the generating
function
\begin{equation}
f_k(z)=\sum_{n=1}^{\infty}\frac{v_n^{(k)}}{\sqrt{n}}z^n=\frac{1}{k}(1-e^{-k\tan^{-1}(z)}),
\end{equation}
with $f_{k=0}(z)=\lim_{k \neq 0,k \rightarrow 0}f_k(z)$.

We introduce the notation
\begin{equation}
|z\rangle=(z,z^2,z^3,\cdots)^{T}, \ |k\rangle=(v^{(k)}_1,v^{(k)}_2,v^{(k)}_3,\cdots)^{T}.
\end{equation}
$\langle z |=|z\rangle^T$ is the transpose of $\langle z |$,
not the hermitian conjugate.
In this notation, the generating function for
the eigenvector $|k\rangle$ can be written as
\begin{equation}
f_k(z)=\langle z|E^{-1}|k\rangle=\langle k|E^{-1}|z\rangle,
\end{equation}
where $E_{nm}=\sqrt{n}\delta_{nm}$.
The crucial properties of $|k\rangle$ are \cite{Okuyama-kinetic}
\begin{eqnarray}
&&\langle k|p\rangle={\cal N}(k)\delta(k-p),\\
&&{\bf 1}=\int_{-\infty}^{\infty}dk
{\cal N}(k)^{-1}|k\rangle\langle k|,\label{completeness}
\end{eqnarray}
where
\begin{equation}\label{N(k)}
{\cal N}(k)=\frac{2}{k}\sinh\left(\frac{\pi}{2}k\right).
\end{equation}

The twist acts as
\begin{equation}
C|z\rangle=|-z\rangle.
\end{equation}
There is a useful relation
\begin{equation}\label{zdz}
z\partial_z\langle z|=\langle z |E^2,
\end{equation}
which leads to
\begin{equation}\label{zEk}
\langle z|E|k\rangle=\langle k|E|z\rangle=\frac{z}{1+z^2}e^{-k\tan^{-1}(z)}.
\end{equation}
The matter solution by Kostelecky and Potting is \cite{Kostelecky-Potting}
\begin{equation}
|\Psi_m \rangle = \exp\left (-\frac{1}{2}\sum_{n,m=1}^{\infty}
a_{n\mu}^{\dagger} S_{nm} a_{m}^{\mu\dagger} \right)|0\rangle.
\end{equation}
The matrix $T\equiv CS$ is given by \cite{Kostelecky-Potting}
\begin{equation}
T=\frac{1}{2M}\left(1+M-\sqrt{(1-M)(1+3M)}\right).
\end{equation}
The eigenvalues of $M$ and $T$ for the eigenvector $|k\rangle$ are known:
\begin{eqnarray}
&&M(k)=-\frac{1}{1+2\cosh(\frac{\pi}{2}k)},\label{M(k)}\\
&&T(k)=-e^{-\frac{\pi}{2}|k|}.\label{T(k)}
\end{eqnarray}

\section{Analytical Proofs}\label{proof}
\setcounter{equation}{0}

\subsection{Ghost Solution}
It is convenient to begin with the ghost solution as we will see later.
The ghost solution by Hata and Kawano is given as \cite{Hata-Kawano}
\begin{equation}
|\Psi_g \rangle = \exp\left (\sum_{n,m=1}^{\infty}
 c_{-n} \widetilde{S}_{nm}b_{-m}\right )c_1|0\rangle,
\end{equation}
where $\widetilde{S}$ can be written in terms of $T$ \cite{Okuyama-kinetic}:
\begin{eqnarray}\label{S-widetilde-CETE}
\widetilde{S}=-CETE^{-1}.
\end{eqnarray}
On the other hand, the sliver in the twisted ghost CFT is defined by the correlation
function \cite{GRSZstructure}
\begin{equation}
\langle \Xi'_g |\phi \rangle \equiv \langle f \circ \phi'(0)
\rangle ',
\end{equation}
where $\phi$ is an arbitrary Fock state and $ f(\xi) = \tan^{-1}(\xi)$.
Since the conformal transformation $f$ acts on the vacuum state
as the exponential of a linear combination of the Virasoro generators,
we have for the sliver in the twisted ghost CFT,
\begin{equation}
|\Xi'_g \rangle = \exp\left (\sum_{n,m=1}^{\infty}
 c_{-n} \widehat{S}_{nm}b_{-m}\right )|0'\rangle,
\end{equation}
The $SL(2,{\bf R})$-invariant vacuum $|0'\rangle$ of the twisted CFT is
$c_1|0\rangle$ of the untwisted CFT.
We will prove the equality of the two states by showing that
the generating functions for $\widehat{S}$ and $\widetilde{S}$ are
the same.

To compute the generating function $\langle z|\widetilde{S}|w\rangle$,
let us consider the function \cite{GRSZstructure}
\begin{eqnarray}
h(z,w) &\equiv& \langle\Xi'_g|R( c'(w)b'(z))c_0|0'\rangle\nonumber\\
&=&\langle 0'|\exp(-\sum_{n,m=1}^{\infty}(-1)^{n+m}c_n
\widehat{S}_{nm}b_m)R( c'(w)b'(z))c_0|0'\rangle \label{h-def-operator}\\
&=& \langle f\circ c'(w) f\circ b'(z) f\circ c'(0) \rangle'.
\label{h-def-cft}
\end{eqnarray}
Here $R$ denotes radial ordering and the extra signs in
(\ref{h-def-operator}) come from BPZ conjugation. The expression
(\ref{h-def-operator}) allows us to write this function as
\footnote{In \cite{GRSZstructure}, the second term of
(\ref{h-operator-computation}) was missing. This term arises from
contraction of $c'(w)$, $b'(z)$, and $c_0$ and is identical with
$\langle c'(w)b'(z)c'(0)\rangle '$.}
\begin{equation}\label{h-operator-computation}
h(z,w)=\sum_{n,m=1}^{\infty} (-w)^m (-z)^{n-1}\widehat{S}_{nm}
+\frac{w}{z(w-z)}.
\end{equation}
Another expression for $h(z,w)$ can be obtained using (\ref{h-def-cft}):
\begin{eqnarray}\label{h-cft-computation}
h(z,w)&=&\langle c'(f(w))b'(f(z))\frac{d f(z)}{dz} c'(f(0))\rangle' \nonumber\\
&=&\frac{1}{1+z^2}\frac{1}{\tan^{-1}(w)-\tan^{-1}(z)}
\frac{\tan^{-1}(w)}{\tan^{-1}(z)}.
\end{eqnarray}
Note that the $c$ ghost has weight zero and the $b$ ghost has weight one in
the twisted ghost CFT.
Combining eqs.(\ref{h-operator-computation}) and (\ref{h-cft-computation}),
we obtain the generating function of $\widehat{S}$.
\begin{eqnarray}
\langle -z|\widehat{S}|-w\rangle &=&
\frac{z}{1+z^2}\frac{1}{\tan^{-1}(z)-\tan^{-1}(w)}
\frac{\tan^{-1}(w)}{\tan^{-1}(z)} + \frac{w}{w-z} \nonumber\\
&=&\langle z|\widehat{S}|w\rangle\label{S-hat-generating-function}.
\end{eqnarray}

Now we compute $\langle z|\widetilde{S}|w\rangle$.
Because of the twist matrix $C$ in the definition of $S$,
it is simpler to use $-z$ instead of $z$.
\begin{eqnarray}
\langle -z|\widetilde{S}|w\rangle&=&
-\int_{-\infty}^{\infty}dk\langle z|ET|k\rangle  N(k)^{-1}
\langle k|E^{-1}|w\rangle \nonumber \\
&=&-\int_{-\infty}^{\infty}dk\frac{k}{2}\frac{1}{\sinh(\frac{\pi}{2}k)}
(-e^{-\frac{\pi}{2}|k|})
\frac{z}{1+z^2}e^{-k\tan^{-1}(z)} \nonumber\\
& &\times
\frac{1}{k}(1-e^{-k\tan^{-1}(w)}) \nonumber\\
&=&\frac{z}{1+z^2}\int_{0}^{\infty}dk
\frac{e^{-\frac{\pi}{2}|k|}}{\sinh(\frac{\pi}{2}k)}
\left( \sinh((\theta+\phi)k) -\sinh(\theta k)\right)
\end{eqnarray}
Here $\theta\equiv \tan^{-1}(z)$, $\phi \equiv \tan^{-1}(w)$,
and we used eqs. (\ref{completeness}), (\ref{N(k)}), (\ref{zEk}),
and (\ref{T(k)}).
We now use the formula
\begin{equation}\label{good-formula}
\int_0^{\infty}dk e^{-\frac{\pi}{2}k}\frac{\sinh(\theta k)}{\sinh(\frac{\pi}{2}k)}
=\frac{1}{\theta} -\frac{1}{\tan\theta},~
(|{\rm Re} ~\theta|<\pi)
\end{equation}
to obtain
\begin{eqnarray}\label{-z-widetilde-S}
\langle -z|\widetilde{S}|w\rangle &=&
-\frac{z}{1+z^2}\frac{\tan^{-1}(w)}{\tan^{-1}(z)(\tan^{-1}(z)+\tan^{-1}(w))}+\frac{w}{z+w}.
\end{eqnarray}
If we replace $z$ by $-z$, this expression exactly coincides with
eq.(\ref{S-hat-generating-function}).
This completes the proof of the equality $\Psi_g=\Xi'_g$.
\subsection{Matter Solution}

The solution by Kostelecky and Potting is given in the operator
language \cite{Kostelecky-Potting}:
\begin{equation}
|\Psi_m \rangle = \exp\left (-\frac{1}{2}\sum_{n,m=1}^{\infty}
a_{n\mu}^{\dagger} S_{nm} a_{m}^{\mu\dagger} \right)|0\rangle.
\end{equation}
We have for the matter sliver \cite{RSZclassical}
\begin{equation}
|\Xi_m \rangle = \exp\left (-\frac{1}{2}\sum_{n,m=1}^{\infty}
a_{n\mu}^{\dagger} S'_{nm} a_{m}^{\mu\dagger} \right)|0\rangle,
\end{equation}
with the coefficient matrix $S'$ determined by the defining equation
\cite{RSZboundary}
\begin{equation}
\langle \Xi_m |\phi \rangle \equiv \langle f \circ \phi(0)
\rangle,
\end{equation}
where $\phi$ is any Fock state and $f(\xi) = \tan^{-1}(\xi)$.
Proceeding in the same way as for the ghost solution,
we will prove the equality of these two states by computing
the generating functions of $S$ and $S'$.

To obtain the generating function of $S'$,
consider the tensor function
\begin{eqnarray}
h^{\mu\nu}(z,w) &\equiv&
\langle \Xi_m|R\left(\partial X^{\mu}(w)\partial X^{\nu}(z)\right)|0\rangle \nonumber \\
&=&\langle 0|\exp\left(-\frac{1}{2}\sum_{n,m=1}^{\infty}(-1)^{n+m}
a_{n\mu}^{\dagger} S'_{nm} a_{m}^{\mu\dagger}\right)
R\left(\partial X^{\mu}(w)\partial X^{\nu}(z)\right)| 0\rangle
\label{h-mu-nu-oscillator}\\
&=&\langle f\circ\partial X^{\mu}(w) f\circ\partial X^{\nu}(z)\rangle.
\label{h-mu-nu-cft}
\end{eqnarray}
The signs in (\ref{h-mu-nu-oscillator}) arise because of BPZ
conjugation. From the expression (\ref{h-mu-nu-oscillator}), we
get
\begin{equation} \label{g-operator-computation}
h^{\mu\nu}(z,w)=
\frac{1}{2}\sum_{n,m=1}^{\infty}(-w)^{n-1}(-z)^{m-1}\sqrt{nm}
S'^{nm}\eta^{\mu\nu}
-\frac{1}{2}\eta^{\mu\nu}\frac{1}{(z-w)^2}.
\end{equation}
The other expression (\ref{h-mu-nu-cft}) allows us
to compute $h^{\mu\nu}(z,w)$ as
\begin{eqnarray} \label{g-cft-computation}
h^{\mu\nu}(z,w) &=&\frac{df(w)}{dw}\frac{df(z)}{dz}
\langle \partial X^{\mu}(f(w))\partial X^{\nu}(f(z))\rangle  \nonumber \\
&=&
\frac{1}{1+w^2}\frac{1}{1+z^2}\left(-\frac{1}{2}\right)\eta^{\mu\nu}
\frac{1}{(\tan^{-1}(z)-\tan^{-1}(w))^2}.
\end{eqnarray}
Comparing eqs.(\ref{g-operator-computation}) and (\ref{g-cft-computation}) gives the
generating function
\begin{eqnarray}\label{S-prime-generating-function}
\langle -z|ES'E|-w\rangle&=&\frac{zw}{(z-w)^2}-
\frac{zw}{(1+z^2)(1+w^2)}\frac{1}{(\tan^{-1}(z)-\tan^{-1}(w))^2}\nonumber\\
&=&\langle z|ES'E|w\rangle.
\end{eqnarray}

The generating function $\langle z|ESE|w\rangle$ for $S$ can be
computed from the generating function for the ghost solution.
\begin{eqnarray}
\langle -z|ESE|w \rangle &=&
\langle z|ETE|w \rangle \nonumber\\
&=&
w\frac{\partial}{\partial w}\langle z|ETE^{-1}|w \rangle\nonumber\\
&=&
-w\frac{\partial}{\partial w}\langle -z|\widetilde{S}|w \rangle\nonumber\\
&=&
-w\frac{\partial}{\partial w}\left(
-\frac{z}{1+z^2}\frac{\tan^{-1}(w)}{\tan^{-1}(z)(\tan^{-1}(z)+\tan^{-1}(w))}+\frac{w}{z+w}
\right)\nonumber\\
&=&\frac{zw}{(1+z^2)(1+w^2)}\frac{1}{(\tan^{-1}(z)+\tan^{-1}(w))^2}
-\frac{zw}{(z+w)^2}.\label{-z-ESE}
\end{eqnarray}
Here we used eqs. (\ref{zdz}), (\ref{S-widetilde-CETE}), and (\ref{-z-widetilde-S}).
Recovering the normal sign, we see that eq.(\ref{-z-ESE}) precisely agrees with
eq.(\ref{S-prime-generating-function}).
This shows $S=S'$ and completes the proof for the matter solution.


\subsection{Candidate State for the tachyon}

The Hata-Kawano state, which is a candidate state for the tachyon, 
is given by \cite{Hata-Kawano}
\begin{equation}
\exp\left(-\frac{1}{2}\sum_{n,m=1}^{\infty}
a^{\dagger}_n S_{nm} a^{\dagger}_m
-a_0 \sum_{n=1}^{\infty} t_n a^{\dagger}_n\right)|k\rangle
\end{equation}
where
\begin{equation}\label{t-def}
(t_n)= |t\rangle
=-\frac{(1-T)^2}{(1-M)(1+T)}|v_{+0}+v_{-0}\rangle.
\end{equation}
The vector $v_{+0}+v_{-0}$ is related to $v_e$ which was
introduced in \cite{Okuyama-tension} through
\begin{equation}
|v_{+0}+v_{-0}\rangle=2|v_{e}\rangle.
\end{equation}
In \cite{Okuyama-tension}, it was found that
\begin{equation}\label{kve}
    \langle k|v_e\rangle=\frac{1}{k}\frac{\cosh\left(\frac{\pi
    k}{2}\right)-1}{2\cosh\left(\frac{\pi
    k}{2}\right)+1}.
\end{equation}
 The matrix factor in (\ref{t-def}) can also
be simplified:
\begin{equation}
\frac{(1-T)^2}{(1-M)(1+T)}=\frac{T^2-T+1}{1+T}.
\end{equation}
Therefore,
\begin{eqnarray}
|t\rangle&=&
-2\frac{T^2-T+1}{1+T}\int_{-\infty}^{\infty}
dk|k\rangle\frac{k}{2\sinh(\frac{\pi}{2})} \frac{1}{k}
\frac{\cosh(\frac{\pi}{2}k)-1}{2\cosh(\frac{\pi}{2}k)+1}\\
&=&
-\frac{1}{2}\int_{-\infty}^{\infty}dk|k\rangle
\frac{T^2(k)-T(k)+1}{1+T(k)}
{(1+T(k))^2\over T^2(k)-T(k)+1}{1\over \sinh({\pi\over 2}k)}\\
&=&
-{1\over 2}\int_{-\infty}^{\infty}dk|k\rangle{(1-e^{-{\pi\over 2}|k|})\over \sinh({\pi\over2}k)}.
\end{eqnarray}
Now let us consider the generating function $\langle z|E|t\rangle$.
\begin{eqnarray}
\langle z|E|t\rangle &=&
-{1\over 2}\int_{-\infty}^{\infty}dk \frac{z}{1+z^2}e^{-k\tan^{-1}(z)}
{(1-e^{-{\pi\over 2}|k|})\over \sinh({\pi\over2}k)}\\
&=&
\frac{z}{1+z^2}\int_0^{\infty}dk(1-e^{-\frac{\pi}{2}k})
\frac{\sinh(\theta k)}{\sinh(\frac{\pi}{2}k)}.
\end{eqnarray}
Here, $\theta=\tan^{-1}(z)$.
Using the formula (\ref{good-formula}) and
\begin{equation}
\int_0^{\infty}dk\frac{\sinh(\theta k)}{\sinh(\frac{\pi}{2}k)}
=\tan\theta,(|{\rm Re}\ \theta|<\frac{\pi}{2}),
\end{equation}
we obtain
\begin{equation}\label{zet}
\langle z|E|t\rangle
=1-\frac{z}{1+z^2}\frac{1}{\tan^{-1}(z)}.
\end{equation}

Rastelli, Sen, and Zwiebach proposed that the Hata-Kawano state is
given by
\begin{equation}\label{tachyon-def}
\langle \chi_T(k)|\psi\rangle\equiv {\cal
N}\lim_{n\rightarrow\infty}n^{2k^2}\langle e^{ik\cdot
X(\frac{n\pi}{4})} f\circ\psi(0)\rangle_{C_n},
\end{equation}
where ${\cal N}$ is a normalization constant and $C_n$ is the upper 
half plane with the identification $z \simeq z+\frac{n}{2}\pi$ 
\cite{RSZboundary}.
Here we will prove their proposal.
 The insertion of the
tachyon vertex operator produces a linear term in the exponential:
\begin{equation}
|\chi_T(k)\rangle=\exp\left(-\frac{1}{2}\sum_{n,m=1}^{\infty}
a^{\dagger}_n S'_{nm} a^{\dagger}_m
-a_0 \sum_{n=1}^{\infty}\widehat{t}_n a^{\dagger}_n\right)|k\rangle
\end{equation}
The coefficient matrix $S'$ is the one for the matter sliver, for
the quadratic term in the exponential represents the action of the
same conformal transformation $f$ on the vacuum. This can also be
directly checked by computing a two-point correlation function. To
compute the coefficient vector $\widehat{t}$, let us consider the
function
\begin{eqnarray}
\langle\chi_T(k)|\partial X^{\mu}(z)|k'\rangle
&=&
\langle k|\exp\left(-\frac{1}{2}a\cdot S'\cdot a
-a_0 \widehat{t}\cdot C\cdot a\right)\partial X^{\mu}(z)|k'\rangle \\
&=& -ik'^{\mu}\langle k|k'\rangle \frac{1}{z}
\left(1-\sum_{m=1}^{\infty}(-1)^m\sqrt{m}z^m\widehat{t}^m\right).
\label{h-mu-operator}
\end{eqnarray}
Here we have used
\begin{eqnarray}
&&\partial X^{\mu}(z)=\frac{p^{\mu}}{iz}+\frac{1}{\sqrt{2}i}\sum_{m\neq 0}
\frac{\alpha_m^{\mu}}{z^{m+1}},\\
&&a_0=a_0^{\dagger}=\sqrt{2}p.
\end{eqnarray}
On the other hand, from definition (\ref{tachyon-def}), we have
\begin{eqnarray}
\langle\chi_T(k)|\partial X^{\mu}(z)|k'\rangle
&=&{\cal N}\lim_{n\rightarrow\infty}n^{2k^2}\langle
e^{ik\cdot X(\frac{n\pi}{4})} f\circ\partial X^{\mu}(z)
 f\circ e^{ik'\cdot X(0)}\rangle_{C_n}\\
&\propto&
\lim_{w\rightarrow\infty}\langle e^{ik\cdot X(w)}
f\circ\partial X^{\mu}(z)f\circ e^{ik'\cdot X(0)}\rangle_{UHP} \\
&\propto&
\frac{-ik'^{\mu}}{1+z^2}\frac{1}{\tan^{-1}(z)}\langle k|k'\rangle.\label{h-mu-cft}
\end{eqnarray}
By comparing (\ref{h-mu-operator}) and (\ref{h-mu-cft}) and
choosing ${\cal N}$ so that both have the same residue for the pole at $z=0$,
we get
\begin{equation}
\langle \widehat{t}|E|-z\rangle
=1-\frac{z}{1+z^2}\frac{1}{\tan^{-1}(z)}
=\langle \widehat{t}|E|z\rangle
,
\end{equation}
which is identical to eq. (\ref{zet}). This completes the proof
for the candidate tachyon state.
\section*{Acknowledgments}
I would like to thank Yuji Okawa and Hirosi Ooguri for
valuable discussions.
\newpage


\renewcommand{\baselinestretch}{0.87}

\begingroup\raggedright\endgroup
\end{document}